\documentclass[%
 aip,
 amsmath,amssymb,
 reprint,%
]{revtex4-1}

\usepackage{graphicx}
\usepackage{dcolumn}
\usepackage{bm}

\usepackage[utf8]{inputenc}
\usepackage[T1]{fontenc}
\usepackage{mathptmx}
\usepackage{etoolbox}
\usepackage[english]{babel}
\usepackage[explicit]{titlesec}

\titleformat{\section}  
  {\large\bfseries\sffamily}     
  {\thesection}         
  {1em}                
  {#1}      

\renewcommand{\thesection}{\Roman{section}.}

\titleformat{\subsection}
  {\normalsize\bfseries\sffamily}  
  {\thesubsection} 
  {1em}  
  {#1}  
  
\renewcommand{\thesubsection}{\Alph{subsection}.}

\makeatletter
\def\bbl@set@language#1{%
  \edef\languagename{%
    \ifnum\escapechar=\expandafter`\string#1\@empty
    \else\string#1\@empty\fi}%
  \@ifundefined{babel@language@alias@\languagename}{}{%
    \edef\languagename{\@nameuse{babel@language@alias@\languagename}}%
  }%
  \select@language{\languagename}%
  \expandafter\ifx\csname date\languagename\endcsname\relax\else
    \if@filesw
      \protected@write\@auxout{}{\string\select@language{\languagename}}%
      \bbl@for\bbl@tempa\BabelContentsFiles{%
        \addtocontents{\bbl@tempa}{\xstring\select@language{\languagename}}}%
      \bbl@usehooks{write}{}%
    \fi
  \fi}
\newcommand{\DeclareLanguageAlias}[2]{%
  \global\@namedef{babel@language@alias@#1}{#2}%
}
\makeatother

\DeclareLanguageAlias{en}{english}

\makeatletter
\def\@email#1#2{%
 \endgroup
 \patchcmd{\titleblock@produce}
  {\frontmatter@RRAPformat}
  {\frontmatter@RRAPformat{\produce@RRAP{*#1\href{mailto:#2}{#2}}}\frontmatter@RRAPformat}
  {}{}
}%
\makeatother

\addto\captionsenglish{}

\addto\captionsenglish{}

\begin{document}

\preprint{AIP/123-QED}

\title{A Study of Superconducting Behavior in Ruthenium Thin Films}
\author{Bernardo Langa Jr.}
\altaffiliation{Authors made equal contributions to this work}
\affiliation{Department of Physics, University of Maryland, College Park, Maryland 20740, USA}
\affiliation{Department of Physics \& Astronomy, Clemson University, Clemson, South Carolina 29634, USA} 
\author{Brooke Henry}
\altaffiliation{Authors made equal contributions to this work}
\affiliation{Department of Physics \& Astronomy, Clemson University, Clemson, South Carolina 29634, USA}
\author{Ivan Lainez}
\affiliation{Department of Physics, University of Maryland, College Park, Maryland 20740, USA}
\affiliation{Department of Physics \& Astronomy, Clemson University, Clemson, South Carolina 29634, USA}
\author{Richard Haight}
\affiliation{IBM T. J. Watson Research Center, Yorktown Heights, New York 10598, USA}
\author{Kasra Sardashti}
\altaffiliation{Corresponding author: ksardash@umd.edu}
\affiliation{Department of Physics, University of Maryland, College Park, Maryland 20740, USA}
\affiliation{Laboratory for Physical Sciences, University of Maryland, College Park, Maryland 20740 USA}

\date{\today}

\begin{abstract}
Ruthenium (Ru) is a promising candidate for next-generation electronic interconnects due to its low resistivity, small mean free path, and superior electromigration reliability at nanometer scales. Additionally, Ru exhibits superconductivity below 1 K, with resistance to oxidation, low diffusivity, and a small superconducting gap, making it a potential material for superconducting qubits and Josephson Junctions. Here, we investigate the superconducting behavior of Ru thin films (11.9–108.5 nm thick), observing transition temperatures from 657.9 mK to 557 mK. A weak thickness dependence appears in the thinnest films, followed by a conventional inverse thickness dependence in thicker films. Magnetotransport studies reveal type-II superconductivity in the dirty limit ($\xi$ > > \textit{\textbf{l}}), with coherence lengths ranging from 13.5 nm to 27 nm. Finally, oxidation resistance studies confirm minimal RuO\textsubscript{x} growth after seven weeks of air exposure. These findings provide key insights for integrating Ru into superconducting electronic devices. 
\end{abstract}

\maketitle

\section{\label{sec:level1}Introduction}
Ruthenium (Ru) is a rare transition metal that has been the subject of many recent studies for applications in the electronics industry.\cite{yoon_large_2019, rogers_process_2024, wang_role_2023, breeden_ru_2022, sobolev_thermally_2022, barmak_epitaxial_2020} In particular, Ru has been identified as a promising candidate to replace copper (Cu) interconnects. Unlike Cu, whose electrical properties degrade at small length scales (<10 nm) due to grain boundary and surface scattering effects, Ru displays favorable properties in low dimensions, including low electrical resistivity (\textasciitilde6.5 \textmu$\Omega$.cm), small mean free path (\textasciitilde6.7 nm), and superior electromigration reliability over Cu.\cite{breeden_ru_2022, gall_search_2020, kim_electromigration_2023, wen_ruthenium_2016} In addition, Cu requires a diffusion barrier such as TiN to prevent its migration into surrounding materials, especially porous low-k dielectrics used in the interconnect layers. As interconnects become smaller, this barrier takes up a larger fraction of the cross-sectional area, further increasing electrical resistivity. In contrast, Ru has lower tendency to migrate into its surrounding materials, eliminating the need for diffusion barriers on low-k dielectrics.\cite{wen_ruthenium_2016}

Beyond favorable electrical characteristics at room temperature, Ru may be attractive for applications in quantum electronics because it exhibits superconductivity at millikelvin (mK) temperatures. Superconductivity in bulk Ru crystals was first reported in 1957 with transition temperatures (T\textsubscript{c}) of 470 mK and critical magnetic fields of 4.6 mT.\cite{hulm_superconducting_1957} Further studies in the 1960s reported the absence of the isotope effect in Ru crystals and powders with T\textsubscript{c}’s ranging from 474~–~509~mK.\cite{finnemore_absence_1962, geballe_absence_1961, gibson_critical_1966} The only report on the superconducting characteristics of Ru thin films is limited to samples deposited by DC magnetron sputtering on silicon substrates using thin Ti adhesion layers. The resulting Ru thin films of 13 to 300 nm thickness displayed T\textsubscript{c}’s between 550 – 700 mK.\cite{ilin_thin_2016} However, several key questions about the superconductivity of Ru thin films remain unanswered, including whether and how film thickness and structure influence key superconducting parameters such as the superconducting gap, coherence length, superconducting type, and transport behavior in the clean versus dirty limits.

Ru has potential for applications in superconducting quantum electronics due to its resistance to surface oxidation and low diffusivity. When exposed to atmospheric conditions for up to a week, Ru produces a very thin oxide layer (<1 nm).\cite{milosevic_resistivity_2018} This few-monolayer oxidation makes Ru a promising candidate in two-level system (TLS) reduction schemes, where the primary superconducting thin film is capped with a thin inert layer to prevent oxidation.\cite{bal_systematic_2024, de_ory_low_2024, burnett_analysis_2016, chang_eliminating_2024} Ru could serve as a promising capping layer due to its inherent superconductivity, low diffusivity, and self-limiting oxidation. Additionally, Ru can be utilized as an electrode material in Josephson Junctions (JJs), particularly asymmetric JJs where electrodes of different superconducting gaps suppress quasiparticle tunneling and increase the relaxation time of superconducting qubits.\cite{marchegiani_quasiparticles_2022, zhang_loss_2020, mcewen_resisting_2024, steffen_characterization_2023} Ru’s small superconducting gap ($\Delta$\textasciitilde75 \textmu eV) and low diffusivity make it an ideal electrode material for highly asymmetric JJs with AlO\textsubscript{x}/Al base layers. Integration of Ru into superconducting electronics is possible only in the thin film form factor. Therefore, it is imperative to evaluate the superconducting characteristics of Ru as thin films rather than bulk crystals.

\begin{table*}
\caption{\label{tab:table1}List of samples presented in this study and their respective labels (used throughout this work), and growth parameters including the growth temperature, average measured thickness (\textit{\textbf{d}}), standard deviations of thicknesses, and sheet resistances (\textbf{R\textsubscript{s}} at 300 and 1 K).}
\begin{ruledtabular}
\begin{tabular}{cccccc}
 Label & \textit{d} (nm) & Growth Temperature & St. Dev (nm) & R\textsubscript{s} at 300K ($\Omega/\square$) & R\textsubscript{s} at 1K ($\Omega/\square$)\\ \hline
 A & 5.7 & RT & - & 12.8 & 10.5 \\
 B & 11.9 & RT & 0.9 & 4.8 & 3.7 \\
 C & 30.4 & RT & 0.8 & 1.8 & 1.3 \\
 D & 45.3 & RT & 2.2 & 1.3 & 0.9 \\ 
 E & 58.9 & RT & 3.0 & 1.0 & 0.7 \\
 F & 82.2 & RT & 3.7 & 0.5 & 0.3 \\
 G & 108.5 & RT & 3.5 & 0.5 & 0.4 \\
 H & 55.1 & 100 K & 2.0 & 0.9 & 0.6 \\
\end{tabular}
\end{ruledtabular}
\end{table*}

\begin{figure*}
\includegraphics[scale=0.55]{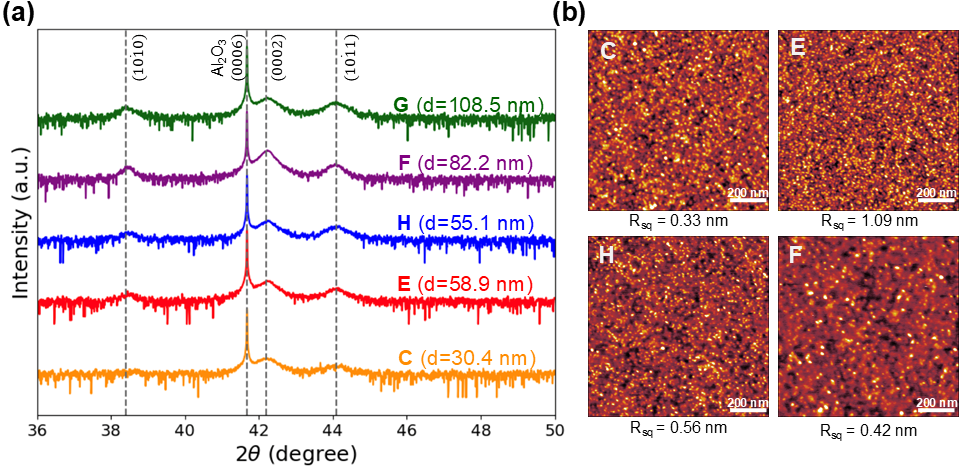}
\caption{\label{fig:FIG1}\textbf{Structural characteristics of the UHV-grown Ru thin films on c-plane sapphire substrates:} (a) X-ray diffraction spectra of thin films with varying thicknesses and growth temperatures (Samples C, E, H, F, and G). (b) AFM topographical maps of Ru thin films of different thicknesses and growth temperatures (Samples C, E, F, and H).}
\end{figure*}

In this study, we investigate the superconducting properties of Ru thin films across a broad thickness range (5.9–108~nm). Ultrahigh vacuum (UHV) electron-beam (e-beam) evaporation was used to grow fine-grained hexagonal Ru films on c-plane sapphire substrates, with root mean square (RMS) roughnesses between 0.33 nm and 1.09 nm. We observe superconductivity in films as thin as 11.9 nm and map the dependence of key parameters–including sheet resistance, critical temperature, critical magnetic field, and coherence length–on film thickness. Using x-ray photoelectron spectroscopy (XPS), we assess the air stability of Ru films over seven weeks of exposure. Our combined materials characterization and magnetotransport studies provide key insights for designing superconducting electronics that leverage the unique physical and chemical properties of Ru thin films.

\section{Results and Discussion}
Using UHV e-beam evaporation, a series of Ru films were grown on c-plane (0001) sapphire substrates with thicknesses (\textbf{\textit{d}}) ranging from 5.7 nm to 108 nm. \textbf{Table \ref{tab:table1}} provides a complete list of samples included in this study. The eight samples are labeled A-H in the first column for ease of referencing throughout the text. During the evaporation, four edges of each chip were covered by a shadow mask. The thickness of each film was directly measured by profilometry across the deposited and undeposited boundaries. Standard deviations of the film thicknesses have also been included in the table. Seven films were grown at temperatures ranging from approximately 25 $^\circ$C to 120 $^\circ$C, referred to as room temperature (RT) for simplicity. The temperature increase resulted from radiation heating due to the electron beam during deposition. One additional film (Sample H) was deposited at 100 K to investigate the impact of cryogenic growth on the film structure and electronic properties. Moreover, another RT-grown Ru film with 64 nm thickness was annealed at 1000 $^\circ$C for 30 minutes to examine the impact of grain growth on the electrical and superconducting characteristics of the Ru thin films.

\begin{figure*}
\includegraphics[scale=0.70]{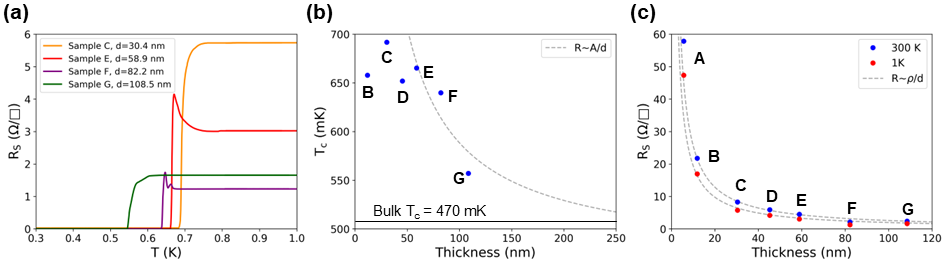}
\caption{\label{fig:FIG2}\textbf{Thickness Dependence of Critical Temperature and Normal Resistance:} (a) Sheet resistance as a function of temperature for Samples C (\textbf{\textit{d}} = 30.4 nm), E (\textbf{\textit{d}} = 58.9 nm), F (\textbf{\textit{d}} = 82.2 nm), and G (\textbf{\textit{d}} = 108.5 nm). (b) Critical temperature as a function of sample thickness fitted to the equation T\textsubscript{c}(\textbf{\textit{d}}) = 470 + \textbf{\textit{A}}/\textbf{\textit{d}}, where \textbf{\textit{A}} = 11.840 K.nm is a fit parameter and \textbf{\textit{d}} is the film thickness. (c) Normal sheet resistance ($\mathbf{R_s}$) vs sample thickness at 300 K and 1 K. The fit uses the equation $R_s=\rho/d$, where $\mathbf{\rho}$ is the resistivity of the film and \textbf{\textit{d}} is the film thickness. The resistivities of the films at 300 K and 1 K were  $\mathbf{\rho_{300K}} = 2.57 \times 10^{-5}$ $\Omega$.cm and $\mathbf{\rho_{1K}} = 1.95 \times 10-5$ $\Omega$.cm.}
\end{figure*}

We examined the crystalline and grain structures of the evaporated Ru films using thin-film X-ray diffractometry (XRD) and atomic force microscopy (AFM). \textbf{Figure \ref{fig:FIG1}(a)} shows the XRD $2\theta$-$\omega$ scans for Samples C, E, F, G, and H. Despite their relatively small thicknesses, films E, H, F and G display three hexagonal Ru peaks measured at 38.4$^\circ$, 42.2$^\circ$, and 44.1$^\circ$ corresponding to the ($10\bar10$), ($0002$), and ($10\bar{1}1$) crystal planes, respectively. Sample C – the thinnest sample studied by XRD with \textbf{\textit{d}} = 30.4 nm – displays only the (0002) and ($10\bar11$) peaks. For each film, a strong peak was also observed at 41.7$^\circ$ corresponding to the c-plane sapphire’s ($0006$) planes. The broad film peaks indicate a fine grain structure of the films. The AFM topography maps for films C, E, H, and F, shown in \textbf{Figure \ref{fig:FIG1}(b)}, confirm the fine-grained nature of the films with RMS roughnesses ranging from 0.33 nm to 1.09 nm. Samples C, E, and H show similar grain sizes while Sample F shows a slight increase in grain size, consistent with its better-resolved XRD spectrum. Moreover, cryogenically-grown Sample H (\textbf{\textit{d}} = 55.1 nm) does not display a significant structural difference compared to Sample E (\textbf{\textit{d}} = 58.9 nm) of nearly identical  thickness. However, the cryogenic growth appears to reduce the RMS roughness from 1.09 nm in Sample E to 0.56 nm in Sample H. Based on the observed XRD spectra, the in-plane and out-of-plane lattice constants of films C-G and the annealed film were calculated to be 2.4 $\mathring {\mathrm A}$ and 3.0 $\mathring {\mathrm A}$, respectively.

Annealing at high temperatures in UHV is expected to promote grain growth in metallic thin films, which aligns with our observation for the 64 nm Ru film annealed at 1000 $^\circ$C for 30 min (see \textbf{Fig. S1} of the \textbf{supplementary material}).\cite{tiggelaar_stability_2009, schmid_effect_2008} The annealing resulted in significant morphological changes, including substantial grain growth, with the largest grains reaching widths of 2 \textmu m. However, the RMS roughness of the annealed sample (1.1 nm) remained nearly identical to that of the unannealed counterpart (Sample E, RMS = 1.09 nm). Additionally, the XRD spectrum showed a marked improvement in clarity, with strong diffraction peaks corresponding to the ($10\bar10$), ($0002$), and ($10\bar11$) crystal planes and distinct Laue oscillations, indicating a highly crystalline film.

The electrical properties of the Ru thin films were analyzed to determine their dependence on film thickness. \textbf{Figure \ref{fig:FIG2}(a)} shows the sheet resistance as a function of temperature for Samples C, E, F, and G, revealing that as thickness increases, T\textsubscript{c} approaches the bulk Ru T\textsubscript{c} of 470 mK. \textbf{Figure \ref{fig:FIG2}(b)} plots T\textsubscript{c} of Samples B, C, D, E, F, and G, defined as the midpoint of the superconducting transition. Sample A (\textbf{\textit{d}} = 5.7 nm) exhibited a metallic behavior down to 270 mK, but did not transition to a superconducting state, suggesting a thickness limit of \textasciitilde10 nm for superconductivity in Ru films. The data was fitted to T\textsubscript{c}(\textbf{\textit{d}}) = 470 + \textbf{\textit{A}}/\textbf{\textit{d}}, where \textbf{\textit{A}} = 11.840 K.nm is a fit parameter dependent on the deposition process and \textbf{\textit{d}} is the film thickness.\cite{marchegiani_quasiparticles_2022, court_energy_2007, cherney_enhancement_1969} Critical temperatures from Samples B, C, and D (\textbf{\textit{d}} = 11.9 - 45.3 nm) were excluded from the fit because they deviated significantly from the expected 1/\textbf{\textit{d}} dependence while an extra point was added for bulk Ru (at the thickness limit of 1 \textmu m). 

\textbf{Table \ref{tab:table2}} presents the critical temperature, residual-resistance ratio (RRR), and Bardeen-Cooper-Schriefer (BCS) superconducting gap ($\Delta\approx $ 1.76\textit{k\textsubscript{B}}T\textsubscript{c}, where \textit{k\textsubscript{B}} is the Boltzmann constant) for each sample. The T\textsubscript{c} values ranged from 557 mK for the thickest film (Sample G) to 657.9 mK for the thinnest superconducting film (Sample B). RRR values increased slightly from 1.28 to 1.71 between Samples B and F, but dropped to 1.46 for Sample G. The superconducting gap followed a similar trend to T\textsubscript{c}, displaying a 1/\textit{\textbf{d}}-like decay for films thicker than 55 nm.\cite{marchegiani_quasiparticles_2022, court_energy_2007} The largest observed $\Delta$ of 104.9~\textmu eV was at 30.4 nm of thickness (Sample C) while the smallest $\Delta$ of 84.5 \textmu eV was seen in the 108.5 nm thick film (Sample G). 

\begin{figure*}
\includegraphics[scale=0.7]{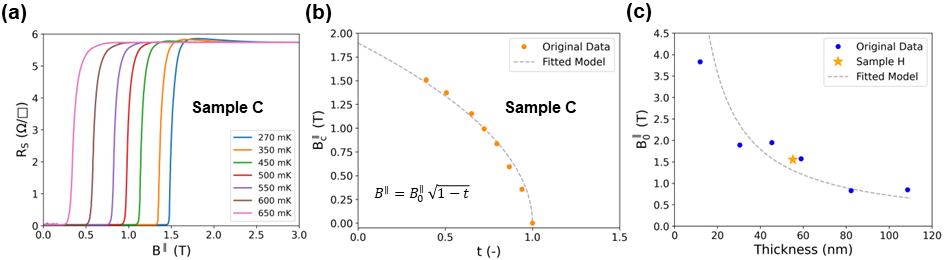}
\caption{\label{fig:FIG3}\textbf{Superconducting-normal transition in parallel magnetic field:} (a) Sheet resistance as a function of the in-plane magnetic field at varying temperatures for Sample C. (b) Critical magnetic field as a function of temperature. The data is fitted using the model: $B_c^\parallel(T)=B_0^\parallel\sqrt{1-t}$. (c) $B_0^\parallel$ values for each Ru thin film fitted to $B_0^\parallel=\sqrt{24}B_{cb}^\parallel(0) \lambda_b(0)/d$.}
\end{figure*}

The T\textsubscript{c} values for films thinner than 50 nm (Samples B, C, and D) deviate from the observed 1/\textbf{\textit{d}} behavior seen for thicker films. Such deviation may be attributed to the films’ thicknesses approaching the 2D limit. Similar behavior has been observed in lead and tin, where increasing the number of monolayers resulted in oscillating T\textsubscript{c} values due to the quantum size effect.\cite{guo_superconductivity_2004, orr_transition-temperature_1984} In films B-F, the number of Ru monolayers ranged between \textasciitilde40-274, respectively. Within this tens to hundreds of atomic monolayer regime, it is possible that the Ru film critical temperatures could be oscillating due to this effect. However, a more precise study is required to verify this behavior and rule out the possibility of impurities formed during the growth process. Beyond 275 monolayers (>82.2 nm), T\textsubscript{c} decreases with increasing film thickness, approximately following a 1/\textbf{\textit{d}} dependence.

\textbf{Figure \ref{fig:FIG2}(c)} shows normal sheet resistance ($\mathbf{R_s}$) at 300 K and 1 K as functions of film thickness for each Ru sample. The exact values are listed in \textbf{Table \ref{tab:table1}}. At both temperatures, $\mathbf{R_s}$ scales with 1/\textbf{\textit{d}} for all film thicknesses studied here. The data fits the conventional metallic behavior with $R_s=\rho/d$, where $\mathbf{\rho}$ is resistivity. Despite the absence of superconductivity, Sample A (\textbf{\textit{d}} = 5.7 nm) exhibited metallic behavior similar to the other films. The resistivities at 300 K and 1 K are $\mathbf{\rho_{300K}}  = 2.57 \times 10^{-5}$ $\Omega$.cm and $\mathbf{\rho_{1K}} = 1.95 \times 10^{-5}$ $\Omega$.cm, respectively. This confirms a consistent conduction mechanism among the films irrespective of their thicknesses. To understand the influence of thin film microstructure on the electrical conduction, we compare the resistivity ($\mathbf{\rho}$) values to the film grown at 100 K (Sample H, \textbf{\textit{d}} = 55.1 nm) where $\mathbf{R_{s,300K}}=3.98$ $\Omega/\square$  and $\mathbf{\rho_{300K}} = 2.19 \times 10^{-5}$ $\Omega$.cm. The relatively small 18\% drop in $\mathbf{\rho_{300K}}$ between the RT and cryogenically-grown sample suggests similar grain structure between the films, which can also be seen in the AFM images [\textbf{Fig. \ref{fig:FIG1}(b)}]. On the other hand, the sample annealed at 1000 $^\circ$C (\textbf{\textit{d}} = 64.2 nm) shows $\mathbf{R_{s,300K}}$ = 1.19 $\Omega/\square$ and a $\mathbf{\rho}$ that is close to the bulk Ru value ($\mathbf{\rho_{300K}} = 7.6 \times 10^{-6}$ $\Omega$.cm).\cite{breeden_ru_2022} The dramatic decrease in $\mathbf{\rho}$ suggests the strong role grain boundaries play in electron scattering within Ru thin films. This agrees with our AFM measurements of the annealed Ru film where larger highly-crystalline grains were observed (see \textbf{Fig. S1} of the \textbf{supplementary material}).

\begin{table*}
\caption{\label{tab:table2}Superconducting properties of Samples B-G, including critical temperature, residual resistance ratio, BCS gap, parallel critical magnetic field, coherence length, and coherence length normalized to the mean free path.}
\begin{ruledtabular}
\begin{tabular}{cccccccc}
 Sample & \textit{d} (nm) & T\textsubscript{c} (mK) & RRR
 & $\Delta$ (\textmu eV) & $B_{0}^{\parallel}$ (T) & $\xi_0$ (nm) & $\xi_0/l$\\ \hline
 B & 11.9 & 657.9 & 1.28 & 0.09 & 3.83 & 27.0 & 13.8 \\
 C & 30.4 & 691.8 & 1.45 & 0.11 & 1.89 & 21.2 & 13.8 \\
 D & 45.3 & 651.9 & 1.42 & 0.09 & 1.95 & 14.2 & 7.28 \\ 
 E & 58.9 & 665.2 & 1.49 & 0.10 & 1.57 & 13.5 & 6.90 \\
 F & 82.2 & 639.9 & 1.71 & 0.09 & 0.83 & 18.1 & 9.28 \\
 G & 108.5 & 557.0 & 1.46 & 0.8 & 0.85 & 13.5 & 6.94 \\
\end{tabular}
\end{ruledtabular}
\end{table*}

\begin{figure*}
\includegraphics[scale=0.7]{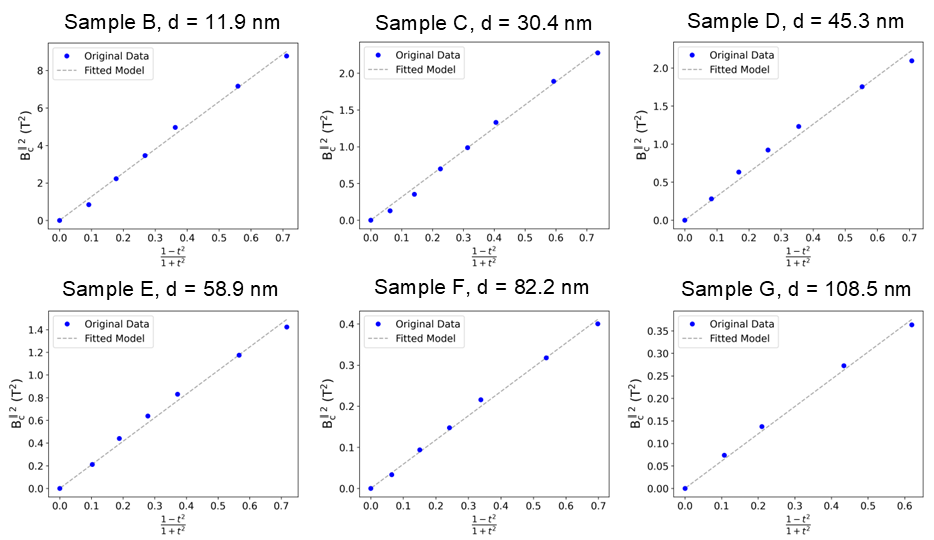}
\caption{\label{fig:FIG4}\textbf{Thickness and temperature dependence of parallel critical magnetic field:} $(B_c^\parallel(T))^2$ as a function of the temperature function $(1-t^2)/(1+t^2)$ for six Ru thin film samples of different thicknesses (Samples B, C, D, E, F, and G). The proportionality factor estimated from the linear fit is $\sqrt{24}\lambda^2(0)(B_c^\parallel(0))^2/d^2$, where $\lambda$ is the London penetration depth and \textbf{\textit{d}} is the thickness of the film.}
\end{figure*}

To further understand the superconducting behavior in Ru thin films, we study the superconducting-normal (S-N) transition of the films in magnetic fields applied parallel to the sapphire substrates. \textbf{Figure \ref{fig:FIG3}(a)} shows the S-N transition for Sample C (\textbf{\textit{d}} = 30.4 nm) in magnetic fields at seven different temperatures from 270 mK and 650 mK. \textbf{Figure \ref{fig:FIG3}(b)} plots Sample C’s critical magnetic fields (at $R_s=0.5R_N$) as a function of normalized temperature t = T/T\textsubscript{c} (T\textsubscript{c} for Sample C is 691.8 mK). To estimate critical magnetic fields at 0 K ($B_0^\parallel$), the Ginzburg-Landau (GL) model for in-plane critical magnetic fields was fit to the data:
\begin{equation}
    B_c^\parallel = B_0^\parallel\sqrt{1-t}
    \label{eq1}.
\end{equation}
Fitting the data to the in-plane GL model leads to a $B_0^\parallel$ of 1.93 T for Sample C. The temperature dependence of $B_c^\parallel$ data for the other six samples was also fit to \ref{eq1} (see \textbf{Table \ref{tab:table2}}, \textbf{supplemental materials Figs. S2 \& S3}). \textbf{Figure \ref{fig:FIG3}(c)} depicts $B_0^\parallel$ for all seven samples, including Sample H grown at 100 K, plotted with a trendline showing the expected thickness dependence of \textbf{$B_0^\parallel$} per GL theory:\cite{reale_thickness_1974}
\begin{equation}
    B_0^\parallel = \sqrt{24}B_{cb}(0) \lambda_b(0)/d
    \label{eq2},
\end{equation}
where \textbf{$B_{cb}$} is the thermodynamic critical field and \textbf{$\lambda_b$} is the effective London penetration depth. Although there are deviations with the thinnest sample (\textit{\textbf{d}} = 11.9 nm), \textbf{$B_0^\parallel$}  generally follows the 1/\textit{\textbf{d}} dependence expected by GL theory. In principle, this fit should also yield an estimate for $\lambda_b(T)$ of Ru, however, uncertainties in the values of $B_{cb}(T)$ can make such an estimation difficult.  Although the $\lambda_b(T)$ values, estimated using the bulk $B_c(0)$ value reported in \textbf{Ref.13}, exhibit a temperature dependence consistent with a dirty superconductor (diverging at T T\textsubscript{c}), their magnitudes are in the micrometer range ($\lambda_b(0)$ = 2.18 \textmu m, see \textbf{ supplemental information in Fig. S4}). Those values are significantly larger than the penetration depth reported for similar superconducting elements (e.g. 520~$\mathring {\mathrm A}$ in bulk Al and 98~$\mathring {\mathrm A}$ in rhenium thin films).\cite{behroozi_penetration_1971, teknowijoyo_superconducting_2023}

The zero temperature coherence length ($\xi_0$) is commonly determined from extrapolated perpendicular $B_c(0)$. However, given the geometrical restrictions in our experimental setup, we estimate the coherence length from the temperature dependence of the parallel critical magnetic field ($\mathbf{B_c^\parallel}$) using the two-fluid model.\cite{cody_magnetic_1968, stenuit_temperature_2004, pethick_relaxation_1979} The temperature dependence of $\mathbf{B_c^\parallel}$ in the two-fluid model is expressed as following as a function of zero temperature penetration depth ($\mathbf{\lambda(0)}$) and film thickness (\textit{\textbf{d}}):\cite{cody_magnetic_1968}
\begin{equation}
    (B_c^\parallel(T))^2 = [24 \lambda^2(0) (B_c^\parallel(0))^2/d^2] (1-t^2)/(1+t^2)
    \label{eq3}.
\end{equation}
\textbf{Figure \ref{fig:FIG4}} displays the dependence of $(B_c^\parallel)^2$ on the temperature function of the two-fluid model $(1-t^2)/(1+t^2)$ for Samples B-G. The dotted line in each panel represents the linear fitted expressed in Eq. \ref{eq3}. The linear relationship observed in \textbf{Fig. \ref{fig:FIG4}} confirms that Ru is a perfect Type II superconductor, similar to behavior seen in lead thin films.\cite{cody_magnetic_1968} However, the GL parameter $\kappa$ was not calculated due to the large uncertainties in our estimations of London penetration depths. 

\begin{figure*}
\includegraphics[scale=0.7]{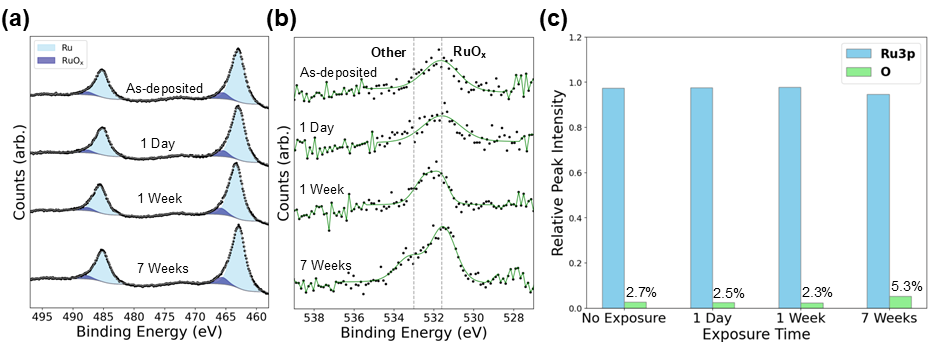}
\caption{\label{fig:FIG5}\textbf{Ru thin film surface composition as a function of air exposure:} (a) Ru3p spectra with metallic Ru and RuO\textsubscript{x} highlighted in light and dark blue, respectively. (b) O1s spectra as a function of exposure time. (c) Surface composition as a function of air exposure calculated from the Ru3p and O1s intensities (normalized to their respective relative sensitivity factor values).}
\end{figure*}

This fit enables the calculation of the product of the thermodynamic critical field ($B_c$) and the effective London penetration depth ($\lambda_{eff}$).\cite{cody_magnetic_1968} From this product, we calculate the GL coherence length at 0 K ($\xi(0)$) using the following relationship:\cite{tinkham_introduction_1996}
\begin{equation}
    \xi(0) = \frac{\Phi_0}{2\sqrt2 \pi B_c^\parallel(0) \lambda(0)}
    \label{eq4}.
\end{equation}
Here, $\Phi_0=2.067\times10^{-15}$ Wb is the magnetic flux quantum. \textbf{Table \ref{tab:table2}} lists the extracted $\xi(0)$ for Samples B-G extracted from measurements in parallel magnetic fields. In general, the coherence length decreases with increasing film thickness, from 27.0 nm in Sample B to 13.5 nm in Sample G. This behavior agrees with the 1/\textbf{d} thickness dependence of $B_c^\parallel$ observed in \textbf{Fig. \ref{fig:FIG3}c}.

We can further determine whether the thin films are in the clean or dirty limit by comparing our estimated $\xi(0)$ values to the electronic mean free path (\textbf{\textit{l}}) in the Ru thin films. In the absence of direct measurements of the electrons’ effective mass, we rely on the $\mathbf{\rho \times}$ \textbf{\textit{l}} products from the literature for Ru thin films ($= 3.81 \times 10^{-16}$ $\Omega$.m\textsuperscript{2} for c-plane and $5.14\times10^{-16}$~$\Omega$.m\textsuperscript{2} for in-plane) to estimate mean free paths.\cite{gall_electron_2016} Considering that two thirds of the Ru peaks in the XRD data (\textbf{Fig. \ref{fig:FIG1}}) have in-plane orientation, the c-plane $\mathbf{\rho \times}$ \textbf{\textit{l}} was chosen as a conservative lower bound. Using the measured resistivity of $\mathbf{\rho_{1K}} = 1.95\times10^{-7}$ $\Omega$.m, the mean free path at 1~K was calculated to be 1.95 nm. The $\xi/l$ ratios calculated in the last column of \textbf{Table \ref{tab:table2}} confirm that all the Ru thin films presented in this study are superconducting in the dirty limit, consistent with Ru’s short bulk electronic mean free path and the films’ nanocrystalline nature.

With superconductivity confirmed in films as thin as 11.9~nm, we investigated the air stability of Ru thin film using X-ray photoelectron spectroscopy (XPS). \textbf{Figure \ref{fig:FIG5}(a)} shows the Ru3p XPS spectra for a 30.4 nm thick Ru film as-deposited and after exposure to air for 1 day, 1 week, and 7 weeks. For each spectrum, metallic Ru peaks (Ru3p\textsubscript{3/2} and Ru3p\textsubscript{1/2}) are highlighted in light blue and Ru oxide (RuO\textsubscript{x}) peaks are highlighted in purple. \textbf{Figure \ref{fig:FIG5}(b)} shows the O1s peaks recorded over the same air exposure timespan. The vertical dotted line found at 531.6 eV represents RuO\textsubscript{x} peaks in the film, while the dotted line at \textasciitilde533 eV represents other oxide peaks (e.g., the metal hydroxides and organic C-O) formed after exposure. The ratio between Ru3p and O1s is calculated in \textbf{Fig. \ref{fig:FIG5}(c)} (normalized to their respective relative sensitivity factors (RSFs)). \textbf{Table \ref{tab:table3}} provides a summary of the ratios between RuO\textsubscript{x} and Ru, the RuO\textsubscript{x} binding energies, and the ratio between O and Ru (corrected by the RSF) after each air exposure step.

The as-deposited film had a minimal air exposure time of approximately 10 minutes during sample preparation. This minimal exposure time did not have a significant effect on the amount of RuO\textsubscript{x} present in the film with the ratio of RuO\textsubscript{x}/Ru only being 0.1 as seen on \textbf{Table \ref{tab:table3}}. This is in agreement with the fs-UPS measurement, where a sharp valence band edge was observed with work function close to the reported values for elemental Ru (see \textbf{supplementary material Fig. S5})\cite{alshareef_temperature_2006, pantisano_ruthenium_2006}. Despite the 7 week exposure, there is no noticeable change in the Ru or the RuO\textsubscript{x} with the ratios staying at 0.09-0.1. The O1s peak intensities, however, grow with the air exposure time. This increase is also noticed in the ratio between Ru3p and O1s where oxygen is relatively stable (~2\%) for the first 3 exposure steps before increasing to 5.3\% at 7-week exposure. Nevertheless, rather than the RuO\textsubscript{x} growth, the increase in the oxygen concentration is attributed to the metal hydroxides and carbonaceous surface contaminants  (containing C–O bonds), contributing the rising shoulder peak at 533 eV.\cite{weidler_x-ray_2017}


\begin{table}[t]
\caption{\label{tab:table3}Summary of the XPS surface analysis results for a Ru thin film (\textit{\textbf{d}} = 30.4 nm) after multiple air exposure steps.}
\begin{ruledtabular}
\begin{tabular}{cccc}
 Air exposure time & RuO\textsubscript{x}/Ru & RuO\textsubscript{x} BE (eV) & [O]/[Ru] \\
\hline
As-deposited & 0.10 & 465.5 & 0.027 \\
1 day & 0.09 & 465.5 & 0.026 \\
1 week & 0.10 & 456.8 & 0.024 \\
7 weeks & 0.09 & 465.6 & 0.056 \\
\end{tabular}
\end{ruledtabular}
\end{table}

\section{Conclusion}

In summary, we observed superconductivity in UHV-evaporated Ru films down to 11.9 nm thickness. Deposition at room and cryogenic temperatures produced polycrystalline films with fine grain structures. Resistivity (on the order of $10^{-5}$ $\Omega$.cm) was found to be thickness-independent and limited by the electron scattering at the grain boundaries. For films thinner than 60 nm, T\textsubscript{c} varied over a small range (651 - 691 mK), while for thicker films, it showed nearly inverse proportionality to thickness. Critical magnetic fields varied inversely with thickness, from $B_0^\parallel$ = 0.85 T (\textit{\textbf{d}} = 108.5 nm) to $B_0^\parallel$ = 3.90 T (\textit{\textbf{d}} = 11.9 nm), corresponding to coherence lengths of 13.5 – 27 nm. All films exhibited Type-II superconducting behavior in the dirty limit. XPS confirmed a stable RuO\textsubscript{x} surface layer, unchanged after seven weeks of air exposure. Our results establish Ru as an air-stable superconducting thin film suitable for integration into superconducting electronics such as Josephson junctions and qubits.

\section{Experimental Section}
\subsection{UHV growth of Ru thin films}
The Ru thin films were deposited using a UHV molecular beam epitaxy (MBE) system with a base pressure below $1\times~10^{-9}$ mbar. The manipulator is equipped with an N\textsubscript{2} cooling line allowing the sample to reach temperatures as low as 100 K during deposition. Before the Ru thin film deposition, $5\times5$ mm\textsuperscript{2} c-plane sapphire (0001) substrates were cleaned by sonication in acetone and isopropyl alcohol followed by N\textsubscript{2} drying. The substrates were then annealed at 150 $^\circ$C in the MBE loadlock under high vacuum ($1\times10^{-7}$ mbar). After loading into the main MBE chamber, substrates were annealed once more at 150 $^\circ$C for 20 minutes to remove any remaining surface contaminants. After cooling the substrates to room temperature, Ru films were grown using e-beam evaporation at rates of 0.5-1.7 $\mathring {\mathrm A}$/s. \textbf{Table \ref{tab:table1}} lists the samples presented in this study including growth temperatures and thicknesses (measured by profilometry).

\subsection{Electrical characterization}
To evaluate the critical temperatures and magnetic fields of the Ru thin films, a cryogenic measurement system (Oxford Instruments Teslatron PT) was used. The system is equipped with a \textsuperscript{3}He probe (base temperature of 260 mK) and a single-axis 12-T superconducting magnet. The differential resistances of the samples as functions of temperature and magnetic field were measured in van der Pauw (VdP) configuration using lock-in amplifiers (NF LI-5650).

\subsection{Structural characterization}
The crystal structures of the thin films were investigated using a Malvern Panalytical Empyrean X-ray diffraction instrument with Cu-K$\alpha$ radiation at 45 kV and 40 mA. Five rocking curves in range $18^\circ \leq 2\theta$-$\omega \leq 25^\circ$ were captured on each film. After a Gaussian filter was applied to the raw data, the five scans were averaged to produce the spectra shown in \textbf{Fig. \ref{fig:FIG1}(a)}. For evaluating the surface topography, an atomic force microscope (AFM, Bruker Nanoscope) was employed. The AFM uses a silicon tip in vibrational mode to scan across the sample surface in areas as large as $90 \times 90$ \textmu m\textsuperscript{2} and as small as $0.5 \times 0.5$ \textmu m\textsuperscript{2}.

\subsection{Photoelectron spectroscopy}
 The XPS data was collected at take-off angle of $45^\circ$ using using a PHI VersaProbe III system with a monochromatic Al-K$\alpha$ X-ray source (hv = 1486.6 eV). The Al anode was operated at 25 W with a voltage of 15 kV. The beam spot size was 100 \textmu m in diameter and the analysis area was $500 \times 500$~\textmu m\textsuperscript{2}. Prior to XPS measurements, the samples were sputtered by Ar+ ions (at 2 kV over a $3 \times 3$ mm\textsuperscript{2} sputtering area) for 12~s to remove most carbonaceous surface contaminants. The sample was stored in an air-clean workstation for air exposure studies. XPS spectra were collected for Ru3s, Ru3p, Ru3d, Ru4p, Ru3p, C1s, N1s, and O1s peaks. To estimate the elemental compositions from the XPS data, the peak intensity for each element was corrected by their respective relative sensitivity factor value.\cite{scofield_theoretical_1973} In addition to XPS, femtosecond Ultraviolet Photoelectron Spectroscopy (fs-UPS) was employed to determine the valence band spectra and the Fermi level positions for two Ru thin films with estimated thickness of 30.4 and 58.9 nm. Fs-UPS uses femtosecond lasers which are split into pump and probe components. Individual harmonics are subsequently directed onto the sample within a UHV analysis chamber which produces a standard UPS Spectra.

 \section{Supplementary Material}
 The \textbf{supplementary material} provides information on: i) the structural and electrical characterization of the Ru thin film sample annealed at 1000 $^\circ$C; ii) estimations of $B_0^\parallel$ for Samples B-G using the Ginzburg-Landau model; iii) magnetotransport characteristics of the cryogenically grown Ru thin film samples; iv) temperature dependence of the bulk Ru London penetration depth; v) valence band spectra of two of the Ru thin film samples measured by femtosecond-UPS.
 

\begin{acknowledgments}
The authors acknowledge the support of the U.S. National Science Foundation (under grant no. 2137776 and grant no. 2329017) and the U.S. Department of Energy (under grant no. DE-SC0023595). B.L. and K.S. acknowledge Kelliann Koehler and the Clemson Electron Microscopy Facility for their assistance conducting the XPS measurements. B.H. acknowledges Chomani Gaspe for her assistance in conducting the XRD measurements.
\end{acknowledgments}

\section*{Data Availability Statement}
 The data that support the findings of this study are available from the corresponding author upon reasonable request.

\bibliographystyle{apsrev4-1}
\bibliography{ru_paper}

\end{document}



\title{Supplemental Information\\
A Study of Superconducting Behavior in Ruthenium Thin Films}

\author{Bernardo Langa Jr.}
\altaffiliation{Authors made equal contributions to this work}
\affiliation{Department of Physics, University of Maryland, College Park, Maryland 20740, USA}
\affiliation{Department of Physics \& Astronomy, Clemson University, Clemson, South Carolina 29634, USA}
\author{Brooke Henry}
\altaffiliation{Authors made equal contributions to this work}
\affiliation{Department of Physics \& Astronomy, Clemson University, Clemson, South Carolina 29634, USA}
\author{Ivan Lainez}
\affiliation{Department of Physics, University of Maryland, College Park, Maryland 20740, USA}
\affiliation{Department of Physics \& Astronomy, Clemson University, Clemson, South Carolina 29634, USA}
\author{Richard Haight}
\affiliation{IBM T. J. Watson Research Center, Yorktown Heights, New York 10598, USA}
\author{Kasra Sardashti}
\altaffiliation{Corresponding author: ksardash@umd.edu}
\affiliation{Department of Physics, University of Maryland, College Park, Maryland 20740, USA}
\affiliation{Laboratory for Physical Sciences, University of Maryland, College Park, Maryland 20740 USA}

\date{\today}

\maketitle 

\begin{figure} [H]
\centering
\includegraphics[scale=0.75]{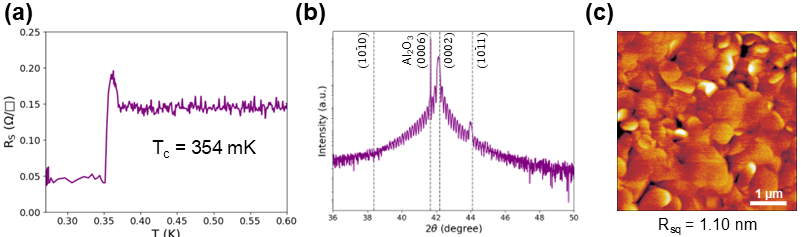}
\caption{\label{fig:FIGS1}\textbf{Critical Temperature and Surface Characterization for a 1000 °C Annealed Ru Thin Film:} (a) Sheet resistance as a function of temperature for the annealed sample with thickness \textit{\textbf{d}} = 64.2 nm. The critical temperature of the sample falls to 354 mK, showing that annealing the film degrades its superconducting properties. (b) XRD spectrum for the annealed Ru thin film displaying distinct Laue oscillations which indicates a highly crystalline film. (c) AFM analysis of the annealed sample reveals the formation of large grains. The normal resistance at 300 K of the sample was 1.19 $\Omega/ \square$, with a resistivity of $\mathbf{\rho} = 7.6 \times 10^{-6}$ $\Omega$.cm.}
\end{figure}

\newpage

\begin{figure} [H]
\centering
\includegraphics[scale=0.65]{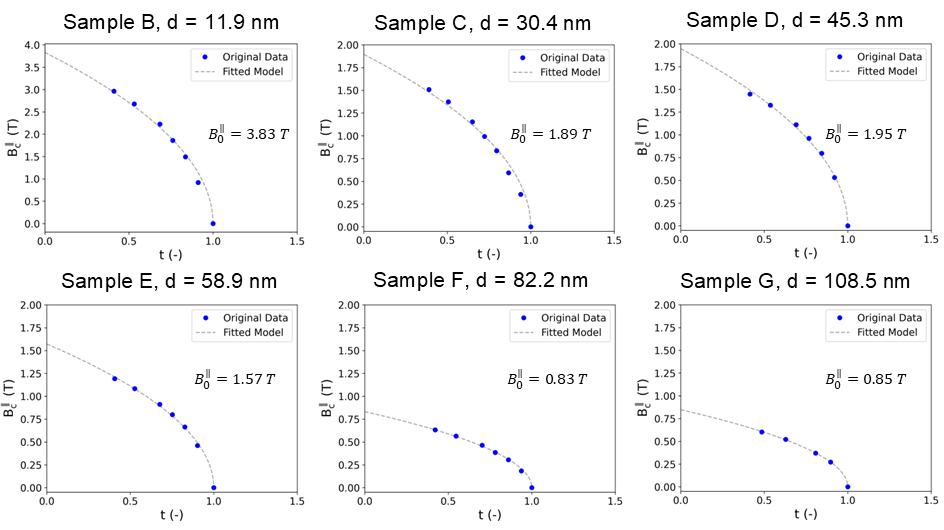}
\caption{\label{fig:FIGS2}\textbf{GL Fit for Estimating of $B_0^\parallel$ of the RT-Grown Ru Films:} Critical magnetic field as a function of  temperature for each RT sample grown. The fit follows the same trend as \textbf{Fig. 3(b)} in the main text. The formula used for the fit is $B_c^\parallel(T)=B_0^\parallel \sqrt{1-t}$, where t = T/T\textsubscript{c}.}
\end{figure}

\newpage

\begin{figure} [H]
\centering
\includegraphics[scale=0.85]{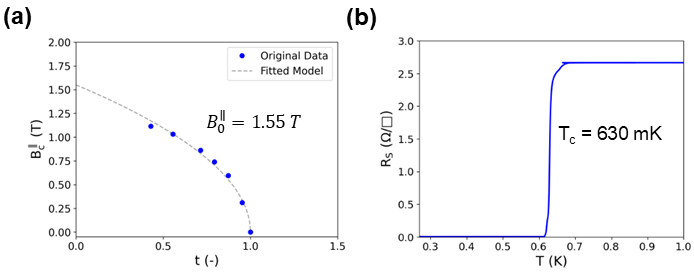}
\caption{\label{fig:FIGS3}\textbf{Magnetotransport characteristics of a Cryogenically-grown Ru Thin Film:} (a) Critical magnetic field as a function of temperature measured for Sample H (grown at 100 K, \textbf{\textit{d}}=55.1 nm). (b) Sheet resistance as a function of temperature showing a critical temperature of 630 mK. The RRR for this sample is 1.48. The normal resistance at 300 K measured in this sample was 4.0 $\Omega/\square$.}
\end{figure}

\newpage

\begin{figure} [H]
\centering
\includegraphics[scale=0.67]{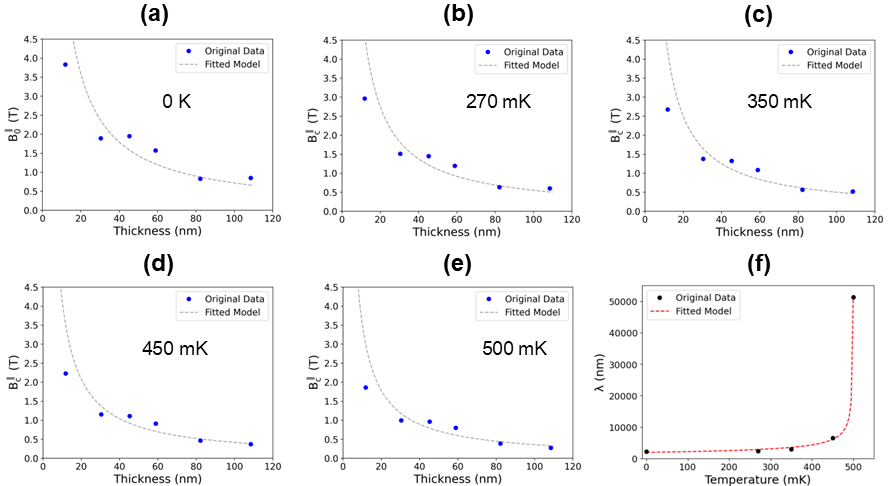}
\caption{\label{fig:FIGS4}\textbf{London Penetration Depth Dependence on Temperature:} (a)-(e) The critical magnetic field was plotted as a function of thickness for five temperatures ranging from 0 to 500 mK. The data was fitted using the model $B_0^\parallel=\sqrt{24}B_{cb}(T)\lambda_b(T)/d$, where \textit{\textbf{d}} represents sample thickness. The thinnest superconducting sample was excluded from the fit due to deviation from the expected trend. (f) Temperature dependence of the bulk London penetration depth ($\lambda$) for Ru obtained from critical parallel magnetic field sweep fits. Divergence of $\lambda$ is observed as the temperature approaches the critical temperature, reflecting the expected behavior near the superconducting transition. The data was fitted using $\lambda_{eff}(T)=\lambda_L(T) \sqrt{\frac{\xi_0}{1.33l}}$, where $\xi_0$ is the coherence length, $l$ is the mean free path,  and $\lambda_L(T)=\frac{\lambda_L(0)}{\sqrt{2(1-t)}}$.}
\end{figure}

\newpage

\begin{figure} [H]
\centering
\includegraphics[scale=0.9]{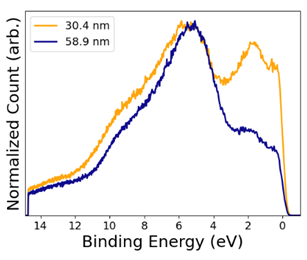}
\caption{\label{fig:FIGS3}\textbf{Valence Band Spectrum of two Ru thin films:} Valence band spectra for two Ru thin film samples with estimated thicknesses of 30.4 nm and 58.9 nm. The thinner sample shows sharper valence bands at the Fermi level edge. The work function for the 30.4 nm film was 4.34 eV, while the 58.9 nm sample had a work function of 4.84 eV, which is well within the range reported for Ru thin films.}

\end{figure}


%
%

%

